\documentclass[preprint,aps,pre,floatfix]{revtex4-1}
\usepackage{amsmath,amssymb}

\usepackage{graphicx}

\begin{document}

\title{Self-consistent Seeding of the Interchange Instability in Dipolarization Fronts}

\author{Giovanni Lapenta (1), Lapo Bettarini (2)}

\affiliation{1) Centrum voor Plasma-Astrofysica, Departement Wiskunde, Katholieke Universiteit Leuven, Celestijnenlaan 200B, 3001 Leuven, Belgium (EU)\\
2) Royal Observatory of Belgium, Ringlaan 3, 1180 Brussels, Belgium (EU)}
\date{\today}
\begin{abstract}
We report a 3D magnetohydrodynamics simulation that studies the formation of dipolarization fronts during magnetotail reconnection.  The crucial new feature uncovered in the present 3D simulation is that the process of reconnection produces flux ropes developing within the reconnection region. These flux ropes are unstable to the kink mode and introduce a spontaneous structure in the dawn-dusk direction. The dipolarization fronts forming downstream of reconnection are strongly affected by the kinking ropes. At the fronts, a density gradient is present with opposite direction to that of the acceleration field and leads to an interchange instability.  We present evidence for a causal link where the perturbations of the kinking flux ropes with their natural and well defined scales drive and select the scales for the interchange mode in the dipolarization fronts. The results of the simulation are validated against measured structures observed by the Themis mission.
\end{abstract}

\maketitle

%
%

%

%
%

\section{Introduction}

Reconnection in the Earth magnetotail is a key step in the processes known as substorms. Reconnection converts magnetic energy stored in the magnetic field of the tail and converts it into heating and plasma flow away from the reconnection site, tailward and Earthward. The Earthward flow is accompanied by a dipolarization front, a front of increased vertical (North-South directed) magnetic field that moves towards the Earth and restores in part a less stretched and more dipole-like magnetic field in the night side of the Earth, thereby acquiring the name of dipolarization front~\citep{nakamura02}. Previous works have investigated the role of the interchange instability in perturbing the dipolarization front~\citep[and references therein]{guzdar10}.

A dipolarization front acts similarly to a snowplow. As the dipolarization front moves, the vertical magnetic field and the density pile up ahead of it.  The curvature of the field lines and the braking of the front by the momentum exchange with the yet unperturbed plasma leads to an effective acceleration pointing contrary to the dipolarization front speed. Such configuration is unstable to interchange modes: the higher density region is ahead of the front and the density gradient is therefore opposite to the direction of the acceleration. Several papers have observed this effect~\citep{pritchett10,saito10,runov09,sitnov09,nakamura02,ge11,birn11}.

Recently, an attempt has been made to match the observed scale of the perturbation of the front with simulations. The published simulations by \citet{guzdar10} are 2D and focus only on the interchange mode, without modeling the full front and its origin from reconnection. By choosing the initial wavelength of the seed, one can control the scale of the evolved structure and reproduce the observed features. In 2D macroscopic (MHD) studies the choice of the initial seed is key because the interchange instability is equally growing in all wavelengths \citep{guzdar10}. In kinetic studies, the finite Larmor radius effect favors the growth rate of smaller wavelength (too small to be of relevance to the observation) \citep{pritchett10}. On the macroscopic scales of the observed structures, MHD models are appropriate and the independence of the growth rate on the wavelength of the modes means that other macroscopic driving forces determine which scales dominate. In 2D studies where only the interchange mode is present, the only way to force the interchange instability is by design.

To avoid the artificial driving due to the selection of the initial seed, a full 3D simulation is needed. We report here a 3D simulation where the dipolarization front and the interchange instability are self-consistently developing out of a reconnection site. We start the system and the reconnection process without seeding any structure along the direction of the  interchange instability. In doing so we let the system develop naturally and form its own desired structures.

As in 2D, the reconnection process in 3D develops a diffusion layer where the plasma and magnetic field decouple and the filed lines reconnect. In the outflow from the reconnection region two regions form where the exhaust from the reconnection region piles up forming dipolarization fronts. Also, as in 2D, the diffusion layer has initially the structure of a Sweet-Parker layer that quickly becomes unstable to the secondary plasmoid instability~\citep{loureiro,lapenta-prl}. 

After that, the progress in 3D becomes radically different from 2D. In 3D, the plasmoid instability leads to the formation of flux ropes ejected with the plasma from the diffusion region and moving towards the dipolarization front. Each flux rope, however, is unstable naturally to the kink instability, a well known instability that has a well defined natural maximum mode of growth identified by the linear mode properties. In our simulations we show that indeed this mode grows to very large amplitudes.  Flux ropes become so distorted as to touch directly the dipolarization fronts, providing a driving to the interchange mode. The dipolarization fronts are naturally unstable to such mode, but the perturbation from the kinking flux rope feeds the interchange instability and determines its dominant mode. 

The results of our simulation predict specific scales for the warping. A direct comparison is possible with observation from the THEMIS mission that can be interpreted as successive crossing of warped fronts. The natural and spontaneous scales developed in our simulations indeed are of the right order to be compatible with the observed scales. 

\section{Macroscopic simulation of reconnection the Earth magnetotail}

We solve the complete set of compressional visco-resistive MHD equations with uniform viscosity and resistivity parametrized by the global Lundquist number, $S$, and kinetic Reynolds number, $R_M$, both set to $10^4$. We use the conventional geomagnetic coordinate system with $x$ in the Sun-Earth direction, $y$ in the dawn-dusk direction and $z$ in the north south direction. The boundary conditions are periodic in $x$ and $y$ and  reflecting (``perfect conductor'') in  $z$.

 The initial configuration has the main field component reversed, $B_x (z) = B_0 \tanh(z/L)$, over a characteristic length scale $L$, while the other components are zero. Reconnection is initiated with a perturbation of the classic GEM-type, repeated identically in every vertical plane, independent of $y$. 

 The system is modeled with FLIP3D-MHD~\citep{flip}, a  Lagrangian Eulerian code based on discretizing the fluid into computational units acting as Lagrangian markers but used to compute the operators on a uniform grid. The simulation reported here used $360$ (in $x$) x $60$ (in $y$) x $120$ (in $z$) Lagrangian markers arrayed initially in a $3$ x $3$ x $3$ uniform formation in each cell of a uniform grid.

The simulation is conducted in dimensionless units, with space normalized to the initial thickness $L$ and time to the Alfv\`en time $\tau_A=v_A/L$.  The computational box has size $L_x/L=60$, $L_y/L=10$ and $L_z/L=10$. In comparing with observations, the results can then be rescaled according to physical units for a chosen observed magnetotail thickness. For example, \citet{guzdar10} report a   initial magnetotail thickness  L=0.2$R_E$ but different thicknesses apply to different conditions of the magnetotail.

\section{Modes of interest}

The evolution displays a sequence of 4 instabilities: tearing, secondary plasmoid, kink and finally interchange instability. Let us follow the evolution. 

\begin{figure}
\centering
\includegraphics[width=.8\columnwidth]{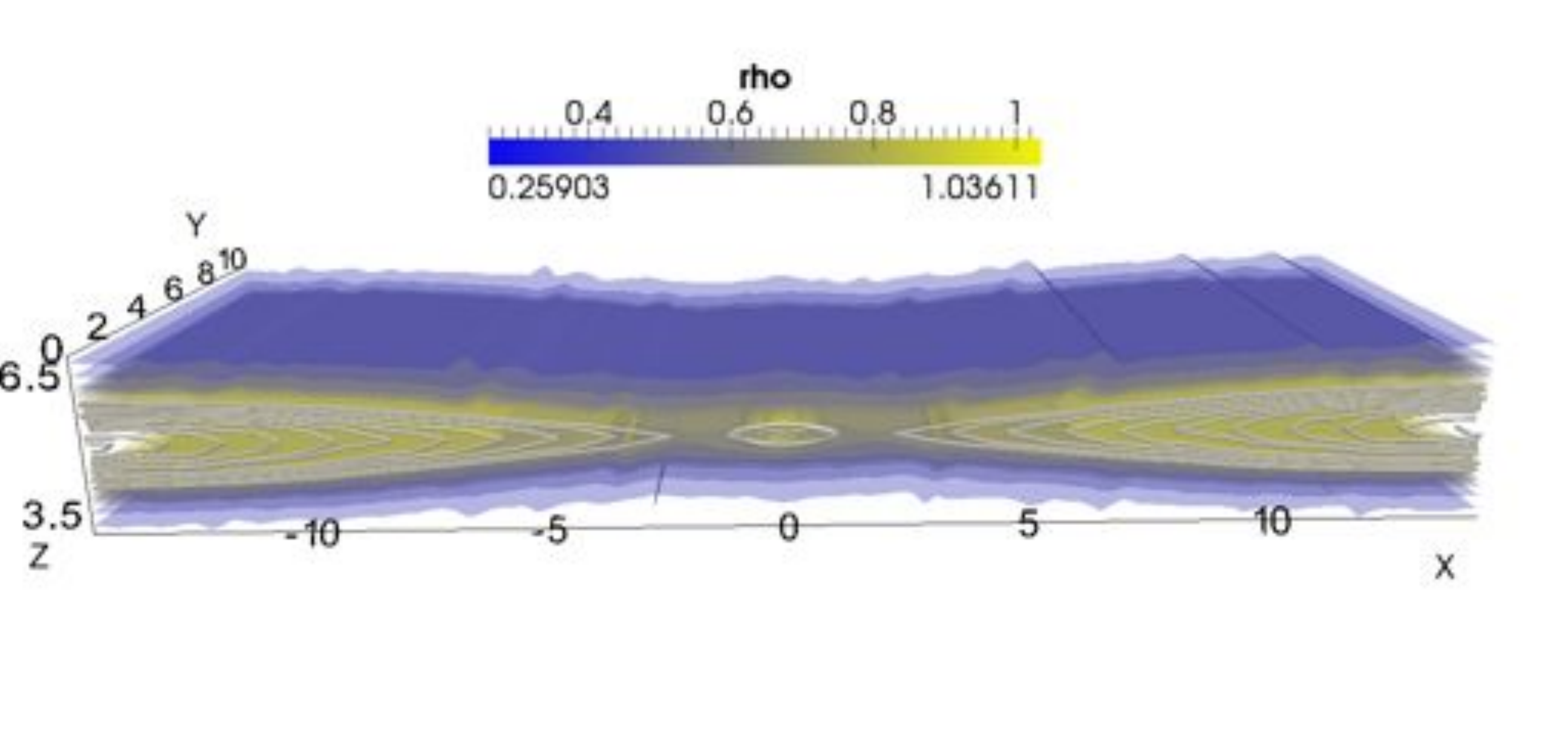}
\caption{3D view of reconnection process. The density structure is shown as nested isosurfaces in color. Selected magnetic field lines are shown. The central reconnection region is harboring  the growth of a secondary island visible in the center as a consequence of the secondary plasmoid instability~\citep{loureiro,lapenta-prl}.}
\label{3d-early}
\end{figure}

 Figure~\ref{3d-early} shows the typical configuration of a reconnection layer forming in response to the initial perturbation localized in the center of the simulation box. From the reconnection layer two jets of plasma exit along the $x$ axis, tailward and Earthward and form the dipolarization fronts:  the areas of curved field lines that resemble dipole-like magnetic field lines. Ahead of the front the plasma piles up, a typical observed feature of dipolarization fronts~\citep[and references therein]{walsh09}. In Fig.~\ref{3d-early} these two regions are visible as areas of increased density (yellow in the color version and light in the grayscale version).

Given the resistivity set for the present simulation, the SP theory leads to a layer of aspect ratio $\Delta/\delta = S^{1/2} = 100 $, being $\Delta$ and $\delta$ the length and the width of the layer respectively. Previous 2D investigations of the same configuration considered here have shown that when the aspect ratio exceeds a theoretical limit (estimated close to 100) a secondary plasmoid instability develops~\citep{loureiro,lapenta-prl}. In the new regime, the process of reconnection proceeds faster and at a rate independent of resistivity. The present 3D simulation is perturbed uniformly in the dawn-dusk direction $y$ and in absence of any other  perturbations the same process develops in each plane and the SP layer becomes unstable leading to plasmoids that in 3D assume the form of flux ropes surrounding current filaments.

For the present investigation the focus is on what happens next. These current filaments have a predominantly azimuthal field, with a small axial component. Therefore, recalling the properties of the kink instability \citep[page 118]{Boyd}, these filaments are violently unstable to the ideal kink instability.

\begin{figure}
\centering
\includegraphics[width=.8\columnwidth]{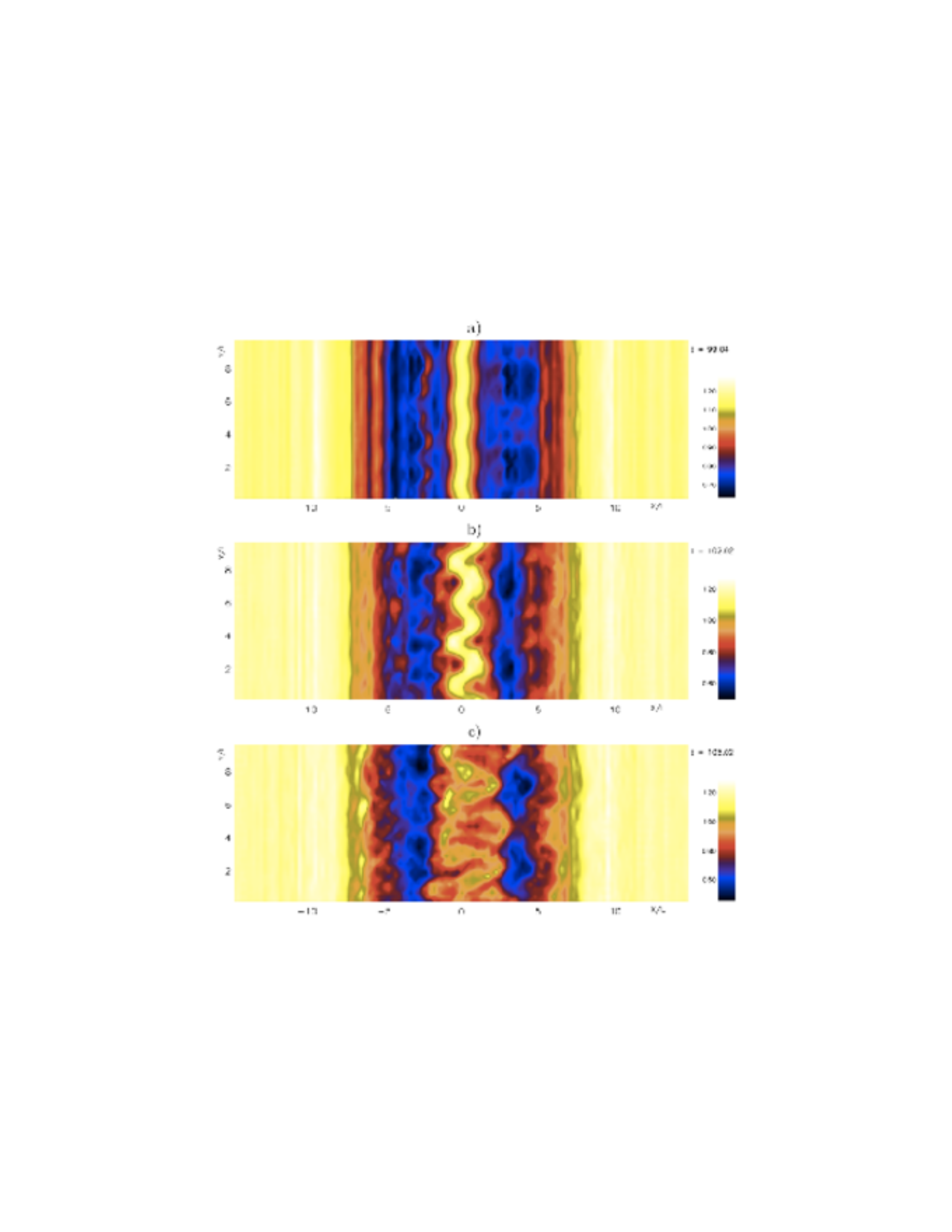}
\caption{Early stage of the evolution. False color representation of the density at 3 different times. Shown are 2D cuts of a 3D run taken on the $xy$ plane at $z=L_z/2$.}
\label{cut-early}
\end{figure}

Figure~\ref{cut-early} provides a view of the evolution of the kink instability seen in a  $xy$ plane cut at $z=0$, the midplane of the simulation box.  

The kink instability develops out of the numerical noise with no prescribed perturbation in the $y$ direction artificially introduced. The noise comes from the truncation error intrinsic to any finite floating point precision in a computer simulation. The present code uses double precision real numbers. This explains the very ordered initial phase of the SP layer formation and its subsequent destabilization by plasmoids. No structure at all is present in $y$ in the initial setup. At first, as the simulation progresses, correctly, the double precision FLIP-MHD code retains perfectly the initial symmetry along $y$, because there is no physical instability that can lead to structure along $y$.

However, as soon as the flux ropes are formed and the stability conditions for the kink instability is exceeded, the very small but present noise from the floating point error seeds the instability breaking the symmetry in $y$ and leading to the kinked rope shown in Fig.~\ref{cut-early}. A dominant mode emerges. 

The evolution of the flux rope is confined within the reconnection region and forms essentially a narrow band with limited span in the vertical ($z$) direction: the flux rope is confined within the reconnecting layer and can be displaced only in the $x$ and $y$ plane. The displacement is then written as 
\begin{equation}
\boldsymbol{\xi}(x,y,z,t) = \widetilde{\boldsymbol{\xi}}(x,z)e^{-i\omega t - i2\pi n y/L_y}
\end{equation}
in terms of eigenfunctions with harmonic dependence in $y$, being $i$ the imaginary constant.

Inspection of Fig.~\ref{cut-early} immediately confirms that the dominant mode has $n=4$. 

The progress of the kink instability leads to a progressive widening of horizontal span of the reconnection region along the $x$ axis affected by the kinked rope. Eventually, the rope starts to come into contact with the dipolarization fronts. Up to this moment the dipolarization front have retained the $y$ invariance of the initial state. In the view of Fig.~\ref{cut-early}, the dipolarization fronts are the two areas of increased density visible at the two ends of the box: left and right. The initial invariance of the simulation along $y$ gives these two regions a perfect straight surface. But as soon as the kink perturbation of the flux ropes becomes significant, another instability develops at the dipolarization fronts and distorts them.

The dipolarization fronts are regions of increasing density where the density gradients points away from the reconnection region. The pile up of density develops in consequence to the motion of the plasma away from the reconnection region, acting as a snow-plow, piling plasma in front of the moving dipolarization front. The present simulation has no gravity but the plasma is still subject to accelerations due to the Lorentz force. As shown in Fig.~\ref{3d-early}, each of the two  dipolarization fronts has a bubble-like structure with a convex surface encasing it. The plasma is inside the convex surface and the field lines circle around it. This  condition is refereed to in the literature as unfavorable because it leads to  instability towards interchange modes. When the field lines at the boundary of the front have a convex shape, the centrifugal force is outwards from the plasma. The density gradient points away from the reconnection region but the acceleration field points towards it. The conditions are met for the interchange instability to develop. In the direction of the acceleration field (pointing toward the reconnection point), the high density region sits above the low density plasma and the instability tends to move the heavy plasma towards the lower energy state, just like a dense liquid on top of a light liquid in a gravitational field would tend to fall down and push the light liquid up (like oil and water).

This fundamental process experienced by everybody  in everyday life (for regular fluids in gravity) leads to the formation of the characteristic pillars (often referred to more ominously as mushrooms) observed in the later stages of Fig.~\ref{cut-early}. 

\begin{figure}
\centering
\includegraphics[width=\columnwidth]{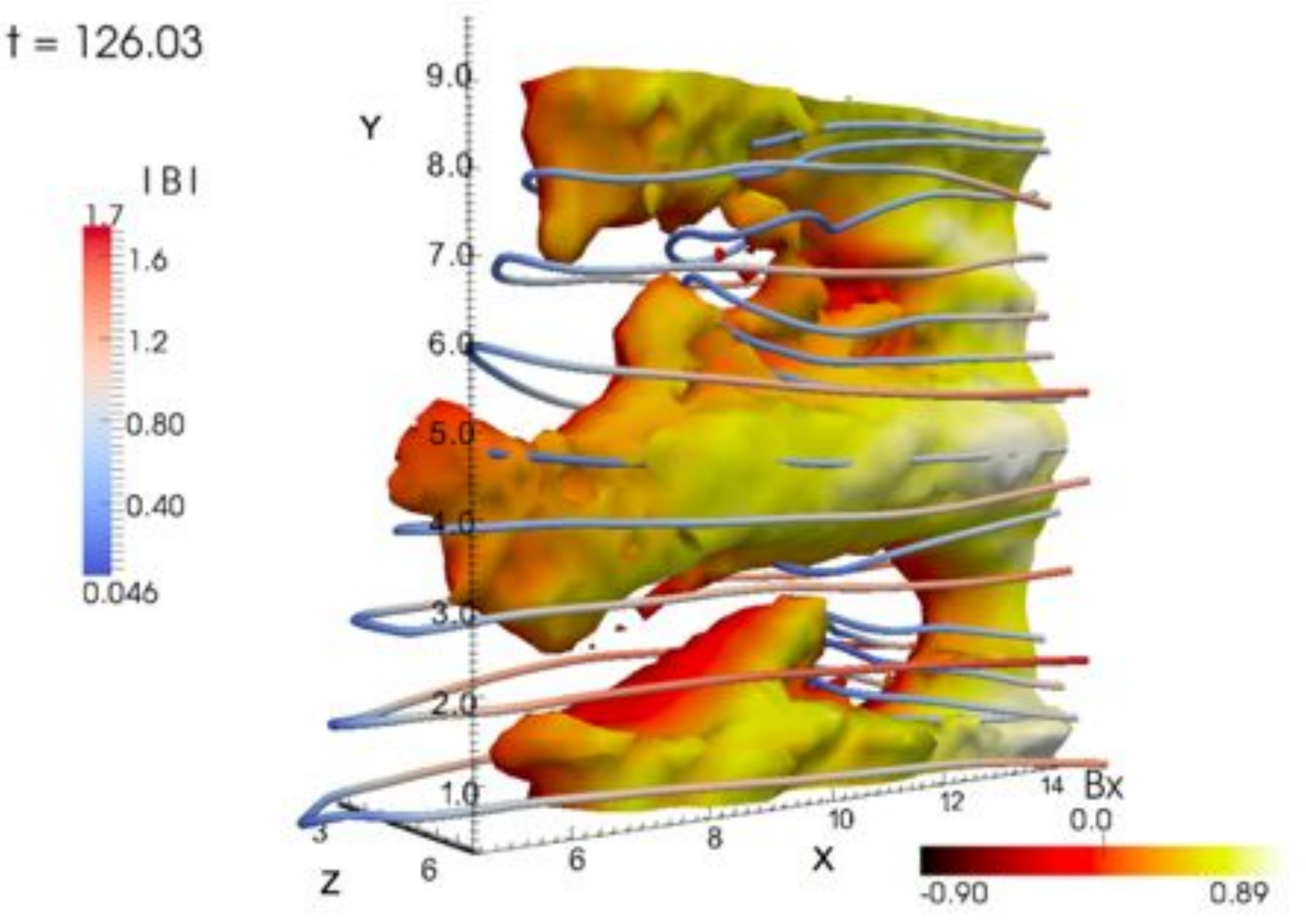}
\caption{3D visualization of the isosurface of density at value 1. Shown is only the right  half of the domain in $x$ at time $t/\tau_A=126.03$ (the same shown in the 2D cut of Fig.~\ref{cut-late}). The surface is colored based on the value of  $B_x$. Field lines are superimposed with their color chosen based on the local intensity of the magnetic field $|\mathbf{B}|$. }
\label{tred}
\end{figure}

\begin{figure}
\centering
\includegraphics[width=\columnwidth]{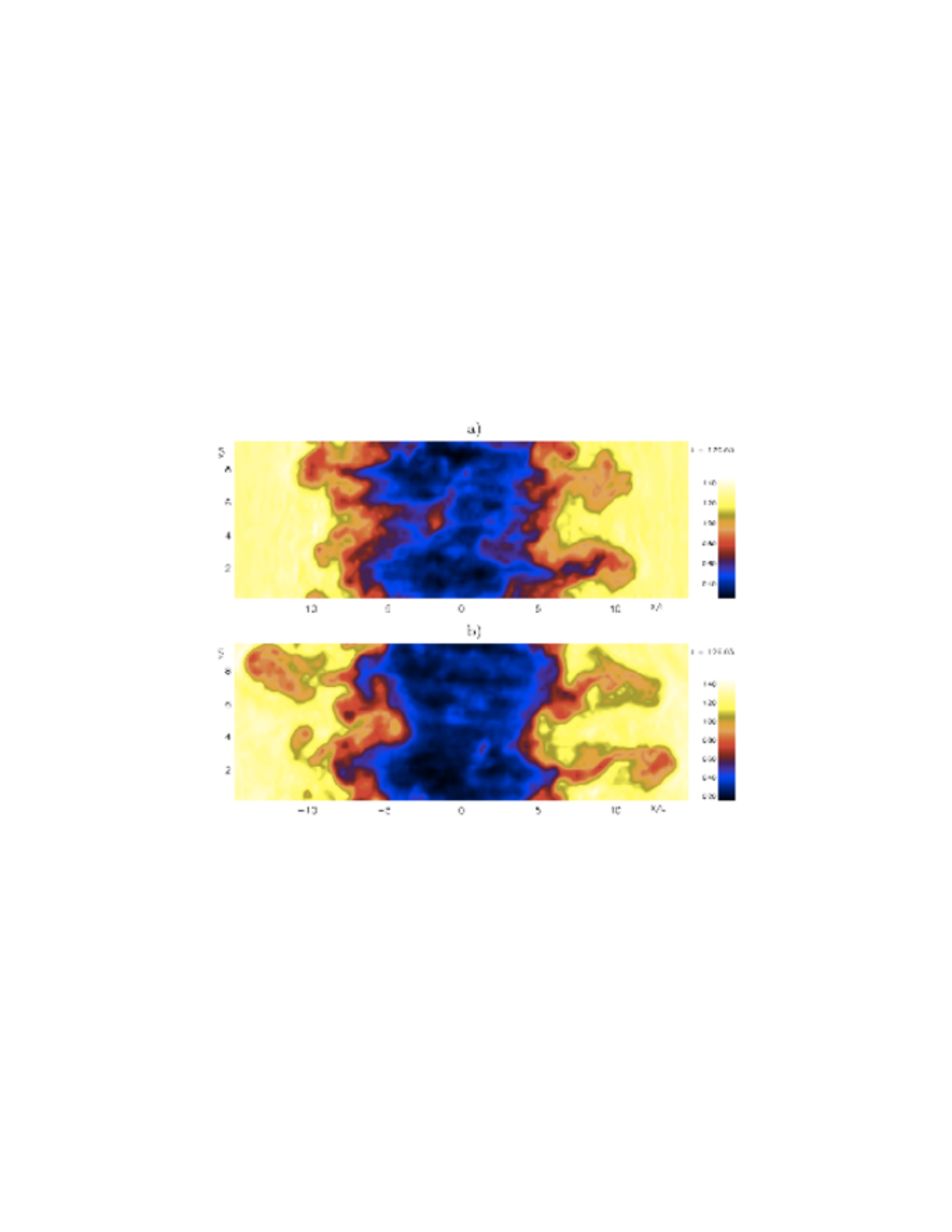}
\caption{Late stage of the evolution. Contour and false color representation of the density at 2 different times. Shown are 2D cuts of a 3D run taken on the $xy$ plane at $z=L_z/2$.}
\label{cut-late}
\end{figure}

The pillars produced by the interchange instability seen in the 2D cuts of Fig.~\ref{cut-early} have a complex 3D shape. Figure~\ref{tred} describes the structure of the pillars on the right side of the simulation box. The dipolarization front is itself a curved front that follows the dipole shape of the field lines. The interchange instability ripples such curved front. Each pillar, that was seen as the characteristic mushroom structure in the 2D cuts, in 3D  wraps around the dipolarization front forming a half-wheel that sticks out of the hub formed by the dipolarization front.  The reconnected field lines  surround and wrap around the front. The pillars produced by the interchange instability continue to stretch and progressively mix leading to a very turbulent state, as shown in the later part of the simulation, reported in Fig.~\ref{cut-late}.

In summary, the  evidence from the simulation is that the instabilities mentioned proceed in strict causal sequence. First the initial perturbation produces a SP reconnecting layer (tearing instability) that becomes unstable to the secondary plasmoid instability. When simulated in 3D the destabilization of the SP layer leads to the formation of flux ropes within the reconnection region. Such flux ropes have primarily an azimuthal magnetic field and are unstable to the ideal kink instability. Concurrently with these events the process of reconnection produces two plasma jets in the outflow and a region of intensified vertical field, the dipolarization fronts. The pile up of the plasma ahead of the dipolarization fronts leads to the conditions for the interchange instability.  The interchange instability of the dipolarization front and the kink instability of the flux ropes produced by the secondary plasmoid instability are both direct consequences of the process of reconnection. Their interplay is considered next.

\section{Model of the kink seeding of the interchange mode}

The conditions for the growth of the interchange instability in the dipolarization front and of the kink instability in the secondary flux ropes develop concurrently and both are the resultant of the modifications produced by reconnection. However, the simulation clearly point to no discernible growth of the interchange mode until the kink of the flux rope is is in the non-linear stage when the flux rope has become visibly distorted, starting to affect the plasma in the vicinity of the dipolarization front. Literally the kink instability touches off the interchange mode and determines the wavelength of the dominant interchange mode.

Both the kink and the interchange instability are ideal modes. However they differ crucially in their spectral properties.

Previous studies of the  interchange instability in MHD models showed that no dominant scale is present and all modes grow with equal rate \citep{guzdar10}. This independence of scale is observed for wavelengths smaller than the length scale of the equilibrium (e.g. \citet[paragraph 19.1]{Goldstone}). This property of the interchange mode is fundamental and shared by many similar models. Even for the gravitational  Rayleigh-Taylor instability, in absence of other effects (dissipations, surface tension) the growth rate is independent of the wavelength (see \citet[ paragraph 91]{Chandrasekhar}) with no fastest growing mode.  This leaves the interchange mode scale neutral, the actual observed scale must therefore be determined by the driving force. If all modes that are excited by the forcing grow with the same rate, it is the intensity of the driving that determines the dominant scales.

The kink instability is, instead, characterized by a well defined fastest growing mode. In the most classical case of a cylindrical flux rope within a concentric cylindrical wall or in vacuum, the well known Kruskal-Shafranov (KS) stability criterion applies: $q(a) = 2 \pi aB_z(a)/L B_\theta(a)>1$  (for a flux rope of radius $a$, periodic length $L$), where $q$ is hte so-called safety factor \citep[page 118]{Boyd}. However, the case under investigation does not correspond to the classical cases of kink of a cylindrical flux rope and the KS limit is only indicative.

 Two main differences are evident. First, the flux rope is confined within the reconnection region that is a relatively narrow band in the $z$ direction. The initial size of this region is equal to the SP layer formed during laminar reconnection. With the evolution of the secondary islands, the region broadens but is still confined by the stronger lobe magnetic field. This confinement in the $z$ direction allows the flux ropes to kink only in the $x-y$ plane and invalidates the usual assumption of cylindrical symmetry of the initial state. 

Second, just like for the secondary plasmoid instability responsible for the breakup of the SP layer and the transition to turbulent reconnection, the presence of flow changes the stability properties quantitatively if not qualitatively~\citep{loureiro}. The flow can stabilize or destabilize the system  as shown in other similar circumstances \citep{Chandrasekhar}. 

For these reasons, we cannot directly compare the instability observed in the simulation with a simple analytical expression for a reference theoretical kink mode. But the simulation indicates a strong predominance for the mode $n=4$. The kinking rope fits four times the domain forming four full sinusoidal oscillations. This has the important implication that  the size of the computational box in the dawn-dusk ($y$) direction is much more than sufficient to capture the fastest growing mode, giving us confidence to be resolving the largest scales of interest. 

The subsequent evolution of the interchange instability demonstrates the initial presence of many modes but in the non-linear later stage the interchange mode $n=2$. This mode is also fully resolved but future work will be needed to follow the expected subsequent interaction of the formed structures. The present simulation in the latest stages suggests that the two pillars formed by interchange tend to interact and further combine into a single structure. Larger simulations can further clarify the later stages of the interchange modes.  

How does this compare with the real magnetotail of the Earth? The observations front the P4 satellite of the THEMIS mission on Feb 15
2008 reported by \citet{guzdar10} show a recurring crossing of the dipolarization from that can be explained assuming that the front itself becomes warped and distorted as a consequence of the interchange instability discussed above. The reported observed scale of such warping is in the range 1-3 $R_E$ confirmed also by previous investigations \citep{nakamura02}. 

The results presented here show a natural and spontaneous emergence of a dominant scale. As noted above the dominant kink and interchange have dominant modes 4 and 2, respectively. The expected length scale of the interchange structures is therefore $L_y/2=5L$. To obtain physical dimensions, we need to choose the normalization length $L$, equal to the reference initial current sheet thickness.  \citet{guzdar10} suggest  L=0.2$R_E$ which leads to a scale of $0.5 R_E$ for the kink mode and $1.0 R_E$ for the interchange mode. This value is only indicative as the magnetotail thickness varies over a wide range. But the agreement with the observations is evident. 

The predicted scale is not determined by the box dimensions. The observed features fit 2 to 4 times the simulations box. This is a clear indication that the scale is determined by the configuration of the initial equilibrium and not by the box size. 
 A key consequence is that this result confirms and extends the conclusions of   \citet{guzdar10} in a  crucial point: in the present case the scales result naturally from the internal nature of the processes developing. Instead, the scales reported by \citet{guzdar10} were intentionally seeded by a suitable choice of the initial perturbation of the front selected by the researchers. 

Future work should investigate via linear theory the conditions of the stability of a flux rope confined within a magnetic field reversal and embedded in a flow pattern representative of the flow present near the reconnection point. This flow has been suggested \citep{loureiro,lapenta-prl} to be a key aspect of the faster turbulent reconnection phase and should be included in a complete model of the evolution. 


%
%
%
%
%
%

%
%
%
%

\begin{acknowledgments}
The present work is supported in part by the NASA MMS mission, by  the Onderzoekfonds KU Leuven (Research Fund KU Leuven) and by the European Commission's Seventh Framework Programme (FP7/2007-2013) under the grant agreement no. 218816 (SOTERIA project, www.soteria-space.eu) and no. 263340 (SWIFF project, www.swiff.eu). The simulations were conducted on the resources of the NASA Advanced Supercomputing Division (NAS), of the The NASA Center for Computational Sciences Division (NCCS) and of the Vlaams Supercomputer Centrum (VSC) at the Katholieke Universiteit Leuven.\end{acknowledgments}

%
%
%
%
%
%
%
%
%
%




%
%


%
%
%
%
%
%



\begin{thebibliography}{}
\bibitem[{\textit{Boyd and Sanderson}(2003)}]{Boyd}
Boyd, T.J.M. and  J.J. Sanderson (2003), The Physics of Plasmas, Cambbridge.

\bibitem[{\textit{Brackbill}(1991)}]{flip}
Brackbill, J. U.  (1991), FLIP MHD: A particle-in-cell method for magnetohydrodynamics, {\it J. Computat. Phys.}, \textit{96}, 163-192.


\bibitem[{\textit{Birn et al.}(2011)}]{birn11}
Birn, J., R. Nakamura, E. V. Panov, and M. Hesse (2011), Bursty bulk flows and dipolarization in MHD simulations of magnetotail reconnection, {\it J. Geophys. Res., 116,} A01210, doi:10.1029/2010JA016083.

\bibitem[{\textit{Chandrasekhar}(1961)}]{Chandrasekhar}
Chandrasekhar, S. (1961), Hydrodynamic and hydromagnetic stability, Oxford: Clarendon.

\bibitem[{\textit{Ge et al.}(2011)}]{ge11}
Ge, Y. S., J. Raeder, V. Angelopoulos, M. L. Gilson, and A. Runov (2011), Interaction of dipolarization fronts within multiple bursty bulk flows in global MHD simulations of a substorm on 27 February 2009,  {\it J. Geophys. Res.,} \textit{116}, A00I23, doi:10.1029/2010JA015758. 

\bibitem[{\textit{Goldston and Rutherford}(1995)}]{Goldstone}
Goldston, R.J.,  P.H. Rutherford (1961), Introduction to plasma physics, Bristol: IOP.

\bibitem[{\textit{Guzdar et al.}(2010)}]{guzdar10}
Guzdar, P.~N., A.~B. Hassam, M. Swisdak, and M.~I. Sitnov (2010), A simple MHD model for the formation of multiple 
dipolarization fronts, {\it Geophys. Res. Lett.,} \textit{37}, L20102, doi:10.1029/2010GL045017. 

\bibitem[{\textit{Lapenta}(2008)}]{lapenta-prl}
Lapenta (2008), Self-Feeding Turbulent Magnetic Reconnection on Macroscopic Scales, {\it Phys. Rev. Lett.}, \textit{100}, 235001.


\bibitem[{\textit{Loureiro et al.}(2007)}]{loureiro}
Loureiro, N. F.,  A. A. Schekochihin, and S. C. Cowley (2007), Instability of current sheets and formation of plasmoid chains, {\it Phys. Plasmas}, \textit{14}, 100703 .

\bibitem[{\textit{Nakamura et al.}(2002)}]{nakamura02}
Nakamura, M.~S., H. Matsumoto, and M. Fujimoto (2002), Interchange instability at the leading part of reconnection jets, {\it Geophys. Res. Lett.,} \textit{29 (N0. 8)}, 1247,doi:10.1029/2001GL013780. 

\bibitem[{\textit{Pritchett and Coroniti}(2010)}]{pritchett10}
Pritchett, P. ~L. and F. V. Coroniti (2010), A kinetic ballooning/interchange instability in the magnetotail, {\it  J. Geophys. Res.,} \textit{115}, A06301,doi:10.1029/2009JA014752. 

\bibitem[{\textit{Runov et al.}(2009)}]{runov09}
Runov, A., V. Angelopoulos, M.~I. Sitnov, V.~A. Sergeev, J. Bonnell, J.~P. McFadden, D. Larson, K.-H. Glassmeier, and U. Auster (2009), THEMIS observation of earthward propagating dipolarization front, {\it Geophys. Res. Lett.,} \textit{36}, L14106, doi:10.1029/2009GL038980. 

\bibitem[{\textit{Saito et al.}(2010)}]{saito10}
Saito, M. ~H., L.-N. Hau, C.-C. Hung, Y.-T. Lai, and Y.-C. Chou (2010), Spatial profile of magnetic field in the near-Earth plasma sheet prior to dipolarization by THEMIS: Feature of minimum B, {\it Geophys. Res. Lett.,} \textit{37}, L08106, doi:10.1029/2010GL042813. 

\bibitem[{\textit{Sitnov et al.}(2009)}]{sitnov09}
Sitnov, M.~I., M. Swisdak, and A.~V. Divin (2009), Dipolarization fronts as a signature of transient reconnection in the magnetotail, {\it J. Geophys. Res.,} \textit{114}, A04202, doi:10.1029/2008JA013980.

\bibitem[{\textit{Walsh et al.}(2009)}]{walsh09}
Walsh, A.~P. et al. (2009), Cluster and Double Star multipoint observations of a plasma bubble, {\it Ann. Geophys.,} \textit{27}, 725-743.






\end{thebibliography}
\end{document}